\documentclass[12pt]{iopart}

\usepackage{iopams}
\usepackage{amssymb,amsfonts,euscript,graphicx,bm}

\usepackage{fancyhdr}            
\fancypagestyle{plain}{
    \lhead{}
    \fancyhead[R]{\thepage}
    \fancyhead[L]{}
    
    \fancyfoot{}
}

\begin{document}





\title{Atomic forces from Dirac-Kohn-Sham equations: Implementation in flexible (APW+lo/LAPW)+LO basis set}

\author{A.~L.~Kutepov}
\address{Condensed Matter Physics and Materials Science Department, Brookhaven National Laboratory, Upton, NY 11973}
\ead{akutepov@bnl.gov}

\begin{abstract}
Atomic forces formulation based on the Dirac-Kohn-Sham equation and flexible (APW+lo/LAPW)+LO basis set is presented. The formulation was implemented in the code FlapwMBPT and allows a user to easily switch between different basis functions of the augmentation type (APW or LAPW) and between different kind of local orbitals. Similar to the work (Phys.Rev.B 91 (2015) 035105), the implementation takes into account small discontinuities of the wave functions, density, and potential at the muffin-tin sphere boundaries. Applications to the materials with strong relativistic effects, such as $\alpha$-Uranium, PuCoGa$_{5}$, and FePt, demonstrate robustness of the method. Comparison of the calculated forces with the ones obtained by numerical differentiation of the free energy shows close agreement with deviations about 0.1\% or less.
\end{abstract}

%
\vspace{2pc}
\submitto{\JPCM}
%
%
%

\section{Introduction}
\label{intr}

One of many achievements of the Density Functional Theory \cite{pr_136_B864,pr_140_A1133} (DFT) is its ability to provide accurate total energy of an interacting many-electron system as a function of external fields such as an electrostatic field generated by nuclei in a solid. Naturally, it is important to be able to find an arrangement of nuclei in a solid (crystal structure) which corresponds to a minimal total energy of the whole system (electrons plus nuclei). Another important field of interest is a studying of the response of a solid when its nuclei are pushed slightly from their equilibrium positions. In both situations, the ability to evaluate accurate derivatives of the total energy with respect to the atomic positions (forces) represents an important tool. In the context of an equilibrium structure search (geometry optimization), availability of the forces greatly helps, providing the directions where atoms should be moved in order to reach their equilibrium positions. In the context of small deviations from the equilibrium (phonons), availability of the forces allows one to use the finite displacement method \cite{epl_32_729,prb_64_045123,prb_78_134106} to calculate phonon frequencies without employing more technically involved linear response approach \cite{rmp_73_515}.

In a family of methods based on the Augmented Plane Waves \cite{pr_51_846} (APW) basis set, such as the Linearized Augmented Plane Waves (LAPW, \cite{prb_12_3060}), the formulation of how one can evaluate atomic forces was given by Yu et al. \cite{prb_43_6411}. The work by Yu et al. demonstrated that breaking the space into non-overlapping muffin-tin (MT) spheres and the so-called interstitial region (IR), which is an important attribute of the APW-family of methods, leads to an additional contribution to the atomic forces, the Pulay term \cite{mp_17_197}. Since then, there were a few enhancements introduced, such as inclusion of additional surface terms when one uses basis sets with discontinuities across the MT boundaries (for instance in APW+lo basis set, \cite{prb_64_195134}). Also, Kl\"{u}ppelberg et al. \cite{prb_91_035105} presented a refinement of the approach by carefully taking into account the tails of the high-energy core states as well as small discontinuities in wafe functions, density, and potential at the MT spheres. Independently, Soler and Williams formulated their variant of the LAPW method \cite{prb_40_1560,prb_42_9728} with perfectly continuous basis functions, as well as the algorithm of the forces evaluation within their approach. Their construction certainly has some advantages, but the complications related to the fact that inside the MT spheres one has to deal with momentum independent functions as well as with momentum dependent plane waves, makes it inconvenient especially if one is interested in advanced methods going beyond DFT such as GW approximation.

One of the limitations of the existing formulations of the atomic force evaluation is that they are based on the non-relativistic Kohn-Sham equations. However, in materials where elements from the far end of the periodic table are present, one has to use fully relativistic approach based on the Dirac-Kohn-Sham (DKS) theory \cite{prb_7_1912,jpc_11_L943,jpc_12_2977,jpc_12_L845}. Therefore, this work has its principal goal in removing the above mentioned limitation. The derivation of the expression for forces goes closely along the lines paved in previous works by Yu et al. and by Kl\"{u}ppelberg et al., but with DKS equations as a background theory. Whereas our derivation is directly relevant to fully relativistic theory, we specifically are pointing out in the text, where the difference from the non-relativistic theory enter. Throughout the paper the atomic units are used with Rydbergs as units of energy.

\section{General derivation of the atomic force expression}
\label{gen}

The force $\mathbf{F}_{t}$ exerted on atom positioned at \textbf{t} is defined as the derivative of the free energy F of a solid: $\mathbf{F}_{t}=-\frac{d F}{d\mathbf{t}}$. Thus, it is convenient to begin with writing down an expression for the free energy which corresponds to a specific level of theory. In the context of a joint description of the relativistic and magnetic effects within the Relativistic Density Functional Theory (RDFT), the corresponding expression was developed in works by Rajagopal, Callaway, Vosko and Ramana \cite{prb_7_1912,jpc_11_L943,jpc_12_2977,jpc_12_L845}. Principal equations of this theory are briefly capitalized here for convenience. In RDFT, free energy of a solid with electronic density
$n(\mathbf{r})$ and magnetization density $\mathbf{m}(\mathbf{r})$ can be written as the following:
\begin{eqnarray}\label{etot}
F &= -T\sum_{\mathbf{k}\lambda}\ln (1+e^{-(\epsilon^{\mathbf{k}}_{\lambda}-\mu)/T})+\mu N 
\nonumber \\&-\int_{\Omega_{0}} d \mathbf{r} [n(\mathbf{r})V_{eff}(\mathbf{r})+\mathbf{m}(\mathbf{r})\cdot
\mathbf{B}_{eff}(\mathbf{r})]\nonumber\\&+\int_{\Omega_{0}} d\mathbf{r} [n(\mathbf{r})V_{ext}(\mathbf{r})+\mathbf{m}(\mathbf{r})\cdot
\mathbf{B}_{ext}(\mathbf{r})] \nonumber\\&+ \int_{\Omega_{0}} d \mathbf{r}
\int_{\Omega} d \mathbf{r'} \frac{n(\mathbf{r})n(\mathbf{r'})}{|\mathbf{r}-\mathbf{r'}|}\nonumber\\&
+\int_{\Omega_{0}} d\mathbf{r} n(\mathbf{r})\epsilon_{xc}[n(\mathbf{r}),\mathbf{m}(\mathbf{r})]+E_{nn},
\end{eqnarray}
where $T$ stands for the temperature, sum runs over the Brillouin zone points $\mathbf{k}$ and band indexes $\lambda$, $\epsilon^{\mathbf{k}}_{\lambda}$ is the band energy, $\mu$ is the chemical potential, and $N$ is the total number of electrons in the unit cell. In the integrals, $\Omega_{0}$ is the volume of the primitive unit cell and $\Omega$ is the volume of the whole solid. 
Effective scalar potential $V_{eff}(\mathbf{r})$ is a sum of an external scalar field $V_{ext}(\mathbf{r})$ and induced fields (Hartree (electrostatic) $V_{H}(\mathbf{r})=2 \int d\mathbf{r'}
\frac{n(\mathbf{r'})}{|\mathbf{r} - \mathbf{r'}|}$ and exchange-correlation $V_{xc}(\mathbf{r})=\frac{\delta
E_{xc} [n(\mathbf{r}),\mathbf{m}(\mathbf{r})]}{\delta n(\mathbf{r})}$):
\begin{equation} \label{veff}
V_{eff}(\mathbf{r})=V_{ext}(\mathbf{r})+V_{H}(\mathbf{r}) + V_{xc}(\mathbf{r}),
\end{equation}
whereas the effective magnetic field $\mathbf{B}_{eff}(\mathbf{r})$ represents a sum of external $\mathbf{B}_{ext}(\mathbf{r})$ and induced $\mathbf{B}_{xc}(\mathbf{r})=\frac{\delta E_{xc}
[n(\mathbf{r}),\mathbf{m}(\mathbf{r})]}{\delta \mathbf{m}(\mathbf{r})}$ magnetic fields:
\begin{equation} \label{beff}
\mathbf{B}_{eff}(\mathbf{r})=\mathbf{B}_{ext}(\mathbf{r})+ \mathbf{B}_{xc}(\mathbf{r}).
\end{equation}

$E_{xc}$ in the above formulae stands for the exchange-correlation energy which is a functional of $n(\mathbf{r})$ and $\mathbf{m}(\mathbf{r})$ : $\int_{\Omega_{0}} d\mathbf{r} n(\mathbf{r})\epsilon_{xc}[n(\mathbf{r}),\mathbf{m}(\mathbf{r})]$. $E_{nn}$ in (\ref{etot}) is the nuclear-nuclear electrostatic interaction energy.
One-electron energies $\epsilon^{\mathbf{k}}_{\lambda}$ are the eigen values of the following equations (Dirac-Kohn-Sham equations):
\begin{equation} \label{ksh}
\left(\hat{K}+V_{eff}(\mathbf{r})+ \beta
\widetilde{\boldsymbol{\sigma}} \cdot \mathbf{B}_{eff}(\mathbf{r})\right) \Psi_{\lambda}^{\mathbf{k}}(\mathbf{r})=\epsilon^{\mathbf{k}}_{\lambda}
\Psi_{\lambda}^{\mathbf{k}}(\mathbf{r}),
\end{equation}
where $\Psi_{\lambda}^{\mathbf{k}}(\mathbf{r})$ stands for the Bloch periodic band function. The kinetic energy operator $\hat{K}$ has the Dirac form (electron rest energy has been subtracted):
\begin{equation} \label{hkin}
\hat{K}=c \boldsymbol{\alpha} \cdot \mathbf{p} +(\beta-I)
\frac{c^2}{2},
\end{equation}
and $\widetilde{\boldsymbol{\sigma}}$ are the $4 \times 4$ matrices, combined from the Pauli matrices $\boldsymbol{\sigma}$:
\begin{equation}
\label{Pauli}
\widetilde{\boldsymbol{\sigma}}= \left(
\begin{array}{cc}
\boldsymbol{\sigma} & 0 \\
0 & \boldsymbol{\sigma}
\end{array} \right).
\end{equation}
$c$ in equation (\ref{hkin}) is the light velocity ($c=274.074$ in our unit system), $\mathbf{p}$ is the momentum operator ($\equiv -i \nabla$),
$\boldsymbol{\alpha}$, and $\beta$ are Dirac matrices in the standard representation, and $I$ is the unit $4 \times 4$ matrix.

Finally, with the electron energies and the band state functions available, the electronic and magnetization densities are defined as the following
\begin{equation} \label{dens}
n(\mathbf{r})=\sum_{\mathbf{k}\lambda}f^{\mathbf{k}}_{\lambda}
\Psi_{\lambda}^{^{\dag}\mathbf{k}}(\mathbf{r}) \Psi_{\lambda}^{\mathbf{k}}(\mathbf{r}),
\end{equation}
and
\begin{equation} \label{magn}
\mathbf{m}(\mathbf{r})=\sum_{\mathbf{k}\lambda}f^{\mathbf{k}}_{\lambda}
\Psi_{\lambda}^{\dag}(\mathbf{k},\mathbf{r}) \beta
\widetilde{\boldsymbol{\sigma}} \Psi_{\lambda}(\mathbf{k},\mathbf{r}),
\end{equation}
with $f^{\mathbf{k}}_{\lambda}$ being the Fermi-Dirac distribution function ($f^{\mathbf{k}}_{\lambda}=\frac{1}{1+e^{(\epsilon^{\mathbf{k}}_{\lambda}-\mu)/T}}$).

Now we are differentiating the Eq.(\ref{etot}) term by term. For the first and the second terms on the right hand side one gets: 

\begin{eqnarray} \label{dif12}
-\frac{d}{d\mathbf{t}}&\left(-T\sum_{\mathbf{k}\lambda}\ln (1+e^{-(\epsilon^{\mathbf{k}}_{\lambda}-\mu)/T})+\mu N\right)
\nonumber\\=&-\sum_{\mathbf{k}\lambda}\frac{1}{1+e^{(\epsilon^{\mathbf{k}}_{\lambda}-\mu)/T}}
\left(\frac{d \epsilon^{\mathbf{k}}_{\lambda}}{d\mathbf{t}}-\frac{d \mu}{d\mathbf{t}}\right)-N\frac{d \mu}{d\mathbf{t}}
=\sum_{\mathbf{k}\lambda}f^{\mathbf{k}}_{\lambda}\frac{d \epsilon^{\mathbf{k}}_{\lambda}}{d\mathbf{t}}.
\end{eqnarray}

The terms from the third to the sixth on the right hand side of (\ref{etot}) are represented by integrals over the unit cell. In APW-related methods, it means technically the sum of the integrals over the non-overlapping MT spheres and over the interstitial region (IR). As authors of work \cite{prb_91_035105} pointed out, the integrals should be differentiated with care, namely, the change of the integration domain when atom (and its muffin-tin sphere) moves should be taken into account. The generic differentiation formula obtained in \cite{prb_91_035105} is the following:

\begin{eqnarray} \label{int_gen}
\frac{d}{d\mathbf{t}}\int_{\Omega_{0}} d\mathbf{r}f(\mathbf{r})=\int_{\Omega_{0}}d\mathbf{r}\frac{d f(\mathbf{r})}{d\mathbf{t}}+\int_{S_{t}}d\mathbf{S}
[f^{MT}(\mathbf{r})-f^{IR}(\mathbf{r})],
\end{eqnarray}
where the surface integral is taken over the MT sphere of atom $t$. $d\mathbf{S}=\mathbf{e}dS$, and $\mathbf{e}=\frac{\mathbf{r}-\mathbf{t}}{|\mathbf{r}-\mathbf{t}|}$ denotes the normal vector on the MT sphere of atom $t$ that points into the interstitial region. $f^{MT}(\mathbf{r})$ and $f^{IR}(\mathbf{r})$ distinguish between the MT and the IR representations of the function $f$. Let us now apply the generic formula (\ref{int_gen}) to the integrals in (\ref{etot}):

\begin{eqnarray} \label{dif3}
-\frac{d}{d\mathbf{t}}&\left(-\int_{\Omega_{0}} d \mathbf{r} [n(\mathbf{r})V_{eff}(\mathbf{r})+\mathbf{m}(\mathbf{r})\cdot
\mathbf{B}_{eff}(\mathbf{r})]\right)\nonumber\\&=\int_{\Omega_{0}} d \mathbf{r} [\frac{d n(\mathbf{r})}{d\mathbf{t}}V_{eff}(\mathbf{r})+n(\mathbf{r})\frac{d V_{eff}(\mathbf{r})}{d\mathbf{t}}\nonumber\\&+\frac{d \mathbf{m}(\mathbf{r})}{d\mathbf{t}}\cdot
\mathbf{B}_{eff}(\mathbf{r})+\mathbf{m}(\mathbf{r})\cdot
\frac{d\mathbf{B}_{xc}(\mathbf{r})}{d\mathbf{t}}]\nonumber\\&+\int_{S_{t}}d\mathbf{S}
[n^{MT}(\mathbf{r})V^{MT}_{eff}(\mathbf{r})-n^{IR}(\mathbf{r})V^{IR}_{eff}(\mathbf{r})]\nonumber\\&+\int_{S_{t}}d\mathbf{S}
[\mathbf{m}^{MT}(\mathbf{r})\cdot \mathbf{B}^{MT}_{eff}(\mathbf{r})-\mathbf{m}^{IR}(\mathbf{r})\cdot \mathbf{B}^{IR}_{eff}(\mathbf{r})],
\end{eqnarray}
where we have assumed that only the induced magnetic field depends of the position of atom $t$.

\begin{eqnarray} \label{dif4}
-\frac{d}{d\mathbf{t}}&\left(\int_{\Omega_{0}} d \mathbf{r} [n(\mathbf{r})V_{ext}(\mathbf{r})+\mathbf{m}(\mathbf{r})\cdot
\mathbf{B}_{ext}(\mathbf{r})]\right)\nonumber\\&=-\int_{\Omega_{0}} d \mathbf{r} [\frac{d n(\mathbf{r})}{d\mathbf{t}}V_{ext}(\mathbf{r})+n(\mathbf{r})\frac{d V_{ext}(\mathbf{r})}{d\mathbf{t}}+\frac{d \mathbf{m}(\mathbf{r})}{d\mathbf{t}}\cdot
\mathbf{B}_{ext}(\mathbf{r})]\nonumber\\&-\int_{S_{t}}d\mathbf{S}
[n^{MT}(\mathbf{r})V^{MT}_{ext}(\mathbf{r})-n^{IR}(\mathbf{r})V^{IR}_{ext}(\mathbf{r})]\nonumber\\&-\int_{S_{t}}d\mathbf{S}
[\mathbf{m}^{MT}(\mathbf{r})\cdot \mathbf{B}^{MT}_{ext}(\mathbf{r})-\mathbf{m}^{IR}(\mathbf{r})\cdot \mathbf{B}^{IR}_{ext}(\mathbf{r})],
\end{eqnarray}

\begin{eqnarray} \label{dif5}
-\frac{d}{d\mathbf{t}}&\left(\int_{\Omega_{0}} d \mathbf{r}
\int_{\Omega} d \mathbf{r'} \frac{n(\mathbf{r})n(\mathbf{r'})}{|\mathbf{r}-\mathbf{r'}|}\right)=-\int_{\Omega_{0}} d \mathbf{r} \frac{d n(\mathbf{r})}{d\mathbf{t}}V_{H}(\mathbf{r})\nonumber\\&-\int_{S_{t}}d\mathbf{S}
[n^{MT}(\mathbf{r})V^{MT}_{H}(\mathbf{r})-n^{IR}(\mathbf{r})V^{IR}_{H}(\mathbf{r})]\nonumber\\&,
\end{eqnarray}

\begin{eqnarray} \label{dif6}
-\frac{d}{d\mathbf{t}}&\left(\int_{\Omega_{0}} d\mathbf{r} n(\mathbf{r})\epsilon_{xc}[n(\mathbf{r}),\mathbf{m}(\mathbf{r})]\right)\nonumber\\&=-\int_{\Omega_{0}} d \mathbf{r} [\frac{d n(\mathbf{r})}{d\mathbf{t}}V_{xc}(\mathbf{r})+\frac{d \mathbf{m}(\mathbf{r})}{d\mathbf{t}}\cdot
\mathbf{B}_{xc}(\mathbf{r})]\nonumber\\&-\int_{S_{t}}d\mathbf{S}
[n^{MT}(\mathbf{r})\epsilon^{MT}_{xc}(\mathbf{r})-n^{IR}(\mathbf{r})\epsilon^{IR}_{xc}(\mathbf{r})],
\end{eqnarray}

Collecting all derivatives together and assuming self-consistency (i.e. equations (\ref{veff}) and (\ref{beff}) are met) we obtain the following force:

\begin{eqnarray} \label{forc1}
\mathbf{F}_{t}=&-\sum_{\mathbf{k}\lambda}f^{\mathbf{k}}_{\lambda}\frac{d \epsilon^{\mathbf{k}}_{\lambda}}{d\mathbf{t}}+\int_{\Omega_{0}} d \mathbf{r} [n(\mathbf{r})\frac{d V_{eff}(\mathbf{r})}{d\mathbf{t}}+\mathbf{m}(\mathbf{r})\cdot
\frac{d\mathbf{B}_{xc}(\mathbf{r})}{d\mathbf{t}}]\nonumber\\&+\int_{S_{t}}d\mathbf{S}
[n^{MT}(\mathbf{r})V^{MT}_{xc}(\mathbf{r})-n^{IR}(\mathbf{r})V^{IR}_{xc}(\mathbf{r})]\nonumber\\&
+\int_{S_{t}}d\mathbf{S}
[\mathbf{m}^{MT}(\mathbf{r})\cdot \mathbf{B}^{MT}_{xc}(\mathbf{r})-\mathbf{m}^{IR}(\mathbf{r})\cdot \mathbf{B}^{IR}_{xc}(\mathbf{r})]\nonumber\\&-\int_{S_{t}}d\mathbf{S}
[n^{MT}(\mathbf{r})\epsilon^{MT}_{xc}(\mathbf{r})-n^{IR}(\mathbf{r})\epsilon^{IR}_{xc}(\mathbf{r})]\nonumber\\&+\mathbf{F}^{HF}_{t},
\end{eqnarray}
where the Hellmann-Feynman force has been introduced:

\begin{equation}\label{hell1}
\mathbf{F}^{HF}_{t}=-\int_{\Omega_{0}} d \mathbf{r} n(\mathbf{r})\frac{d V_{ext}(\mathbf{r})}{d\mathbf{t}}-\frac{d}{d\mathbf{t}} E_{nn}.
\end{equation}

Hellmann-Feynman force is proportional to the gradient of the full electrostatic potential at the center of atom $t$ (excluding the field from its nuclear), \cite{prb_43_6411}:
\begin{eqnarray}\label{hell2}
\mathbf{F}^{HF}_{t}&=2Z_{t}\frac{\partial}{\partial\mathbf{t}} \left(\int_{\Omega} d \mathbf{r} \frac{n(\mathbf{r})}{|\mathbf{t}-\mathbf{r}|}-\sum'_{\mathbf{R}\mathbf{t}'}\frac{Z_{t'}}{|\mathbf{t}-\mathbf{t}'-\mathbf{R}|}\right)\nonumber\\&=-Z_{t}\nabla V'_{el-stat}(\mathbf{r})\bigg|_{\mathbf{r}\rightarrow \mathbf{t}},
\end{eqnarray}
where $Z_{t}$ is the nuclear charge of atom $t$, the integration in the first right hand side expression is performed over the whole solid, and the sum is taken over all unit cells (indexed here by translation vector $\mathbf{R}$) and over all atoms in the unit cell (atom $t$ in the central unit cell is excluded from the sum). This consideration makes the evaluation of the Hellmann-Feynman term easy.

In order to bring the remaining terms of (\ref{forc1}) to the form convenient for evaluation one has to consider the derivative of the one-electron energies. This is done in the next section. Let us also to point out that the  derivation performed up to this point is quite generic with respect to the degree of inclusion of the relativistic effects. The only formal difference is that we use vectors of the magnetization and the magnetic field as it is usually done in the spin-polarized RDFT, instead of spin up and spin down quantities as it is done in the non-relativistic spin-polarized DFT.

\section{Specifics of differentiation of the Dirac-Kohn-Sham eigenvalues}
\label{dkse}

Differentiation of the Kohn-Sham (Dirac-Kohn-Sham) eigenvalues with respect to atomic positions is rather involved. In order to keep derivation as clear as possible we will do it in a step by step fashion. Essentially the derivation is very similar to the one done by Yu et al. \cite{prb_43_6411} and by Kl\"{u}ppelberg et al. \cite{prb_91_035105}. We repeat all the steps here to make it clear where the fully relativistic formalism enters and where the formulae are independent on the formalism (relativistic or non-relativistic). We will consider the derivatives of the valence and core states separately beginning with the valence states.

As a first step, we show explicitly that only the derivatives of the basis functions enter the expression for the forces but not the derivatives of the coefficients. It can be done generically without specification of the basis set or relativistic effects. In methods which use non-orthogonal basis sets the eigenvalues can be found as the ratio of the expectation values of the hamiltonian and overlap matrices:

\begin{equation}\label{eig}
\epsilon=\frac{\sum_{ij}A^{*}_{i}H_{ij}A_{j}}{\sum_{ij}A^{*}_{i}O_{ij}A_{j}},
\end{equation}
where sums run over the basis set indexes and $A_{i}$ are the expansion coefficients. Again, using generic differentiation which we denote as prime, we obtain:

\begin{eqnarray}\label{deig}
\epsilon'&=\sum_{ij}\left(A'^{*}_{i}H_{ij}A_{j}+A^{*}_{i}H'_{ij}A_{j}+A^{*}_{i}H_{ij}A'_{j}\right)\nonumber\\&+\epsilon \sum_{ij}\left(A'^{*}_{i}O_{ij}A_{j}+A^{*}_{i}O'_{ij}A_{j}+A^{*}_{i}O_{ij}A'_{j}\right)\nonumber\\&=\sum_{ij}\left(A'^{*}_{i}[H_{ij}-\epsilon O_{ij}]A_{j}+A^{*}_{i}[H'_{ij}-\epsilon O'_{ij}]A_{j}+A^{*}_{i}[H_{ij}-\epsilon O_{ij}]A'_{j}\right)\nonumber\\&=\sum_{ij}A^{*}_{i}[H'_{ij}-\epsilon O'_{ij}]A_{j},
\end{eqnarray}
where we have used the fact that matrix equations are solved numerically exactly (i.e. for instance $\sum_{j}[H_{ij}-\epsilon O_{ij}]A_{j}$ is zero with computer accuracy). From (\ref{deig}), it is obvious that we have to differentiate only the matrix elements but not the coefficients.

Before proceeding further, let us briefly specify the basis functions (or their combinations) which we are using. As it becomes common practice in the APW-based calculations \cite{cpc_184_2670,cpc_220_230,arx_2012_04992}, we use generic combination of an augmentation function (APW or LAPW) and local orbitals of different kind. As local orbitals, we use the so called 'lo'-orbitals which have discontinuity in its small component (in its derivative in non-relativistic formulation) at the MT sphere boundaries. It is used in combination with APW augmentation \cite{prb_64_195134} to improve variational flexibility of the basis set. Next type of the local orbital is the so called High Derivative Local Orbitals (HDLO) \cite{prb_74_045104,cpc_184_2670,cpc_220_230} which can be used in combination with LAPW or APW+lo to further enhance the accuracy of the basis set in the range of energies corresponding to the valence bands. Finally, the so called High Energy Local Orbitals (HELO's, \cite{cpc_184_2670,cpc_220_230}) can be included in a basis set to describe semicore states or high energy states in the conduction band range of energies.

Hamiltonian and overlap matrix elements are represented by the volume integrals over all MT spheres in the unit cell and over the interstitial region. For the basis functions with discontinuities at the MT surfaces (for instance if the APW+lo combination is used), matrix elements of the hamiltonian include the surface correction terms as it was specified in Ref. \cite{prb_64_195134} for the non-relativistic case and in \cite{arx_2012_04992} for the fully relativistic case. The recipe (\ref{int_gen}) is applied for the differentiation when integration domain changes. Still using generic indexes for the basis set but specifying the band index and the \textbf{k}-point (i.e. using $f^{\mathbf{k}}_{i}$ as generic basis functions and $\epsilon^{\mathbf{k}}_{\lambda}$ as eigenvalues) as well as the specific form (\ref{ksh}) of the Dirac-Kohn-Sham hamiltonian $H_{DKS}$, we obtain:

\begin{eqnarray}\label{dks}
\hspace*{-2cm}
\frac{d H^{\mathbf{k}}_{ij}}{d\mathbf{t}}-\epsilon^{\mathbf{k}}_{\lambda}\frac{d O^{\mathbf{k}}_{ij}}{d\mathbf{t}}&=
P^{\mathbf{k}\lambda}_{ij}\nonumber\\&+\int_{\Omega_{0}} d\mathbf{r}
f^{^{\dagger}\mathbf{k}}_{i}(\mathbf{r})[\frac{d V_{eff}(\mathbf{r})}{d\mathbf{t}}+\beta
\widetilde{\boldsymbol{\sigma}} \cdot \frac{d\mathbf{B}_{eff}(\mathbf{r})}{d\mathbf{t}}]f^{\mathbf{k}}_{j}(\mathbf{r})\nonumber\\&\hspace*{-3cm}+\int_{S_{t}}d\mathbf{S}
\left(f^{^{\dagger}\mathbf{k}(MT)}_{i}(\mathbf{r})[H_{DKS}-\epsilon^{\mathbf{k}}_{\lambda}]f^{\mathbf{k}(MT)}_{j}(\mathbf{r})-f^{^{\dagger}\mathbf{k}(IR)}_{i}(\mathbf{r})[H_{DKS}-\epsilon^{\mathbf{k}}_{\lambda}]f^{\mathbf{k}(IR)}_{j}(\mathbf{r})\right),
\end{eqnarray}
where the terms which later will contribute to the Pulay force have been collected into the quantity $P^{\mathbf{k}\lambda t}_{ij}$:

\begin{eqnarray}\label{pulay1}
P^{\mathbf{k}\lambda t}_{ij}&=\int_{\Omega_{0}} d\mathbf{r}
\frac{d f^{^{\dagger}\mathbf{k}}_{i}(\mathbf{r})}{d\mathbf{t}}[H_{DKS}-\epsilon^{\mathbf{k}}_{\lambda}]f^{\mathbf{k}}_{j}(\mathbf{r})\nonumber\\&+\int_{\Omega_{0}} d\mathbf{r}
f^{^{\dagger}\mathbf{k}}_{i}(\mathbf{r})[H_{DKS}-\epsilon^{\mathbf{k}}_{\lambda}]\frac{d f^{\mathbf{k}}_{j}(\mathbf{r})}{d\mathbf{t}}+\frac{d}{d\mathbf{t}}S^{\mathbf{k}(DISC)}_{ij}.
\end{eqnarray}
Derivatives of the terms which appear in the hamiltonian when some of the basis functions have discontinuities were denoted as $\frac{d}{d\mathbf{t}}S^{\mathbf{k}(DISC)}_{ij}$. We do not specify them here because they will be combined with other explicitly dependent on the atomic position terms in the same way as they were combined in the derivation of the matrix elements of the hamiltonian \cite{arx_2012_04992}.

At this point of the derivation we have to take into account the differences between basis functions of augmentation type (APW and LAPW) and local basis functions (lo, HDLO, and HELO). Also, taking the derivatives assumes an understanding of the quantities themselves. So, in order to avoid the repetition of a rather lengthy derivation of the basis functions and matrix elements which has been done in \cite{arx_2012_04992}, we ask the reader to have the paper \cite{arx_2012_04992} at hand for quick references (we will refer to the equations in that paper as (I-???) with '???' as the equation number). Keeping this in mind, let us proceed with formal differentiation.

For the augmentation functions defined in (I-12,39), the derivative is not zero only in the MT sphere of atom $t$:

\begin{equation} \label{dbasmt}
\frac{d}{d\mathbf{t}}\Pi^{\mathbf{k}}_{\mathbf{G}s}(\mathbf{r})= i(\mathbf{k}+\mathbf{G})\Pi^{\mathbf{k}}_{\mathbf{G}s}(\mathbf{r})-\nabla \Pi^{\mathbf{k}}_{\mathbf{G}s}(\mathbf{r}),
\end{equation}
where the first term comes from the augmentation constraints and the second from the dependence of the radial functions on atomic position. Derivative of the local functions (I-46) also has two terms stemming from a formal Bloch factor and from the same position dependence:
\begin{equation} \label{dbasloc}
\frac{d}{d\mathbf{t}}\Lambda^{\mathbf{k}}_{tnil\mu}(\mathbf{r})= i\mathbf{k}\Lambda^{\mathbf{k}}_{tnil\mu}(\mathbf{r})-\nabla \Lambda^{\mathbf{k}}_{tnil\mu}(\mathbf{r}).
\end{equation}

Let us first consider the contribution of the gradient terms (which is generic) from (\ref{dbasmt}) and (\ref{dbasloc}) into the quantity $P^{\mathbf{k}\lambda t}_{ij}$ in (\ref{pulay1}):

\begin{eqnarray}\label{pulay2}
\hspace*{-2cm}
-\int_{\Omega_{t}} d\mathbf{r}
\nabla &f^{^{\dagger}\mathbf{k}}_{i}(\mathbf{r})[H_{DKS}-\epsilon^{\mathbf{k}}_{\lambda}]f^{\mathbf{k}}_{j}(\mathbf{r})-\int_{\Omega_{t}} d\mathbf{r}
f^{^{\dagger}\mathbf{k}}_{i}(\mathbf{r})[H_{DKS}-\epsilon^{\mathbf{k}}_{\lambda}]\nabla f^{\mathbf{k}}_{j}(\mathbf{r})\nonumber\\&=-\int_{\Omega_{t}} d\mathbf{r}
\nabla \left(f^{^{\dagger}\mathbf{k}}_{i}(\mathbf{r})[\hat{K}-\epsilon^{\mathbf{k}}_{\lambda}]f^{\mathbf{k}}_{j}(\mathbf{r})\right)-\int_{\Omega_{t}} d\mathbf{r}
V_{eff}(\mathbf{r})\nabla [f^{^{\dagger}\mathbf{k}}_{i}(\mathbf{r})f^{\mathbf{k}}_{j}(\mathbf{r})]\nonumber\\&-\int_{\Omega_{t}} d\mathbf{r}
\nabla [f^{^{\dagger}\mathbf{k}}_{i}(\mathbf{r})\beta 
\widetilde{\boldsymbol{\sigma}} \cdot f^{\mathbf{k}}_{j}(\mathbf{r})]\mathbf{B}_{eff}(\mathbf{r})\nonumber\\&=-\int_{S_{t}} \mathbf{e} dS
f^{^{\dagger}\mathbf{k}(MT)}_{i}(\mathbf{r})[\hat{K}-\epsilon^{\mathbf{k}}_{\lambda}]f^{\mathbf{k}(MT)}_{j}(\mathbf{r})-\int_{\Omega_{t}} d\mathbf{r}
V_{eff}(\mathbf{r})\nabla [f^{^{\dagger}\mathbf{k}}_{i}(\mathbf{r})f^{\mathbf{k}}_{j}(\mathbf{r})]\nonumber\\&-\int_{\Omega_{t}} d\mathbf{r}
\nabla [f^{^{\dagger}\mathbf{k}}_{i}(\mathbf{r})\beta 
\widetilde{\boldsymbol{\sigma}} \cdot f^{\mathbf{k}}_{j}(\mathbf{r})]\mathbf{B}_{eff}(\mathbf{r}).
\end{eqnarray}

Let us now consider the contribution from the augmentation parts of the derivatives in (\ref{dbasmt}) and (\ref{dbasloc}). It is easier to take the derivative of the final matrix element, however. In this case one can automatically include the derivatives of the discontinuities because the corresponding contributions to the matrix elements have exactly the same structure of explicit dependence on the atomic positions as the volume integral contributions \cite{arx_2012_04992}. Distinguishing the cases of the matrix elements between two augmentation functions (AA) specified in (I-60,61,68), between the local and the augmentation function (BA) specified in (I-63,64,69), and between two local functions (BB, I-66,67,70) one obtains the corresponding contribution to the quantity (\ref{pulay1}):

\begin{eqnarray}\label{dhmtaa}
\hspace*{-2cm}
i(\mathbf{G}'-\mathbf{G})&\Big[F^{t}_{\mathbf{G'-G}}\sum_{il}D^{\mathbf{k}il}_{\mathbf{G}s;\mathbf{G}'s'}[\overline{h}^{til}_{\mathbf{GG}'}-\epsilon^{\mathbf{k}}_{\lambda}\overline{o}^{til}_{\mathbf{GG}'}]\nonumber\\&+\sum_{il\mu
;i'l'\mu'}\sum_{(ww')=1}^{N^{t}_{l}}y^{^{*}(w)\mathbf{k}}_{til\mu;\mathbf{G}s}
y^{(w')\mathbf{k}}_{ti'l'\mu';\mathbf{G'}s'} \int_{\Omega_{t}}
R^{^{\dagger}(w)t}_{il\mu}(\mathbf{r}) \hat{H}_{NMT}
R^{(w')t}_{i'l'\mu'}(\mathbf{r}) d \mathbf{r}\Big]
\end{eqnarray}
for the AA type, and
\begin{eqnarray}\label{dhmtba}
i\mathbf{G}'\Big[&F^{t}_{\mathbf{G'-G}}\sum_{il}D^{\mathbf{k}il}_{\mathbf{G}s;\mathbf{G}'s'}[\overline{h}^{til}_{\mathbf{GG}'}-\epsilon^{\mathbf{k}}_{\lambda}\overline{o}^{til}_{\mathbf{GG}'}]\nonumber\\&+\sum_{i'l'\mu'}\sum_{w'=1}^{N^{t}_{l}}
y^{(w')\mathbf{k}}_{ti'l'\mu';\mathbf{G'}s'} \int_{\Omega_{t}}
R^{^{\dag}(LOC)t}_{nil\mu}(\mathbf{r}) \hat{H}_{NMT}
R^{(w')t}_{i'l'\mu'}(\mathbf{r}) d \mathbf{r}\Big]
\end{eqnarray}
for the BA type. Derivatives of the matrix elements of BB type equal to zero. The above expressions (\ref{dhmtaa}) and (\ref{dhmtba}) comprise a matrix with indexes running over the whole basis set. Anticipating a convolution of this matrix with the variational coefficients (see Eq. (\ref{deig})), it is convenient to denote this convolution as $\mathbf{C}^{\mathbf{k}t}_{\lambda}$ for a future use. The equations (\ref{dhmtaa}) and (\ref{dhmtba}) are the place where most of the differences between the fully relativistic and the non-relativistic formulations are concentrated. Whereas it is not the goal of this work to give a comprehensive account of all levels of the relativistic effects, it is helpful to know where the differences are located. Particularly, if one needs to recover all non-relativistic equations, the quantities $D^{\mathbf{k}il}_{\mathbf{G}s;\mathbf{G}'s'}$, $\overline{h}^{til}_{\mathbf{GG}'}$, and $\overline{o}^{til}_{\mathbf{GG}'}$ which are defined in (I-59,60,61) for the fully relativistic case, have to be replaced with their non-relativistic analogues.

Now it is a time to perform the Brillouin zone and the band index sums in the basis set convolution of the expression (\ref{dks}) and, correspondingly, to evaluate the first term on the right hand side of (\ref{forc1}):
\begin{eqnarray}\label{dksS}
\hspace*{-2cm}
-\sum_{\mathbf{k}\lambda}f^{\mathbf{k}}_{\lambda}\frac{d \epsilon^{\mathbf{k}}_{\lambda}}{d\mathbf{t}}&=-\sum_{\mathbf{k}\lambda}f^{\mathbf{k}}_{\lambda}\sum_{ij}A^{^{*}\mathbf{k}}_{i\lambda}\Big[\frac{d H^{\mathbf{k}}_{ij}}{d\mathbf{t}}-\epsilon^{\mathbf{k}}_{\lambda}\frac{d O^{\mathbf{k}}_{ij}}{d\mathbf{t}}\Big]A^{\mathbf{k}}_{\lambda}\nonumber\\&=
\mathbf{F}^{Pulay}_{t,val}-\int_{\Omega_{0}} d\mathbf{r}
[n_{val}(\mathbf{r})\frac{d V_{eff}(\mathbf{r})}{d\mathbf{t}}+\mathbf{m}_{val}(\mathbf{r})\beta\widetilde{\boldsymbol{\sigma}} \cdot \frac{d\mathbf{B}_{eff}(\mathbf{r})}{d\mathbf{t}}]\nonumber\\&\hspace*{-3cm}-\sum_{\mathbf{k}\lambda}f^{\mathbf{k}}_{\lambda}\int_{S_{t}}d\mathbf{S}
\Big[\Psi^{^{\dagger}\mathbf{k}(MT)}_{\lambda}(\mathbf{r})[H_{DKS}-\epsilon^{\mathbf{k}}_{\lambda}]\Psi^{\mathbf{k}(MT)}_{\lambda}(\mathbf{r})-\Psi^{^{\dagger}\mathbf{k}(IR)}_{\lambda}(\mathbf{r})[H_{DKS}-\epsilon^{\mathbf{k}}_{\lambda}]\Psi^{\mathbf{k}(IR)}_{\lambda}(\mathbf{r})\Big],
\end{eqnarray}
with the valence Pulay force
\begin{eqnarray}\label{pulay3}
\mathbf{F}^{Pulay}_{t,val}&=-\sum_{\mathbf{k}\lambda}f^{\mathbf{k}}_{\lambda}\mathbf{C}^{\mathbf{k}t}_{\lambda}
+\sum_{\mathbf{k}\lambda}f^{\mathbf{k}}_{\lambda}\int_{S_{t}} d\mathbf{S}
\Psi^{^{\dagger}\mathbf{k}(MT)}_{\lambda}(\mathbf{r})[\hat{K}-\epsilon^{\mathbf{k}}_{\lambda}]\Psi^{\mathbf{k}(MT)}_{\lambda}(\mathbf{r})\nonumber\\&+\int_{\Omega_{t}} d\mathbf{r}
V_{eff}(\mathbf{r})\nabla n_{val}(\mathbf{r})+\int_{\Omega_{t}} d\mathbf{r}
\nabla [\mathbf{m}_{val}(\mathbf{r})]\cdot\mathbf{B}_{eff}(\mathbf{r}).
\end{eqnarray}

For the core states we can formally repeat all above steps which we have done for the valence states, with a number of simplifications. The simplifications are related to the following two facts: i) each core state is an exact solution of the Dirac-Kohn-Sham equation for a spherically symmetric potential as opposite to an expansion in a basis set for the valence levels; ii) core states are strictly confined inside the corresponding MT sphere with zero values and derivatives at the boundary. As a result, all surface terms related to the augmentation or the discontinuities disappear. Equations (\ref{dksS}) and (\ref{pulay3}) for the core states, therefore, can be simplified as the following:
\begin{eqnarray}\label{dksSc}
\hspace*{-2cm}
-\sum_{c}\frac{d \epsilon_{c}}{d\mathbf{t}}&=
\mathbf{F}^{Pulay}_{t,cor}-\int_{\Omega_{0}} d\mathbf{r}
[n_{cor}(\mathbf{r})\frac{d V_{eff}(\mathbf{r})}{d\mathbf{t}}+\mathbf{m}_{cor}(\mathbf{r})\beta\widetilde{\boldsymbol{\sigma}} \cdot \frac{d\mathbf{B}_{eff}(\mathbf{r})}{d\mathbf{t}}],
\end{eqnarray}
with $c$ running over the core states of atom $t$ and with the core Pulay force
\begin{eqnarray}\label{pulay3c}
\mathbf{F}^{Pulay}_{t,cor}=\int_{\Omega_{t}} d\mathbf{r}
V_{eff}(\mathbf{r})\nabla n_{cor}(\mathbf{r})+\int_{\Omega_{t}} d\mathbf{r}
\nabla [\mathbf{m}_{cor}(\mathbf{r})]\cdot\mathbf{B}_{eff}(\mathbf{r}).
\end{eqnarray}

Finally we can include the contribution from the eigenvalue derivatives (\ref{dksS}) and (\ref{dksSc}) into a general force equation (\ref{forc1}) to finish the derivation:

\begin{eqnarray} \label{forc2}
\mathbf{F}_{t}=\mathbf{F}^{HF}_{t}+\mathbf{F}^{Pulay}_{t,cor}+\mathbf{F}^{Pulay}_{t,val}+\mathbf{F}^{Surf}_{t,kin}+\mathbf{F}^{Surf}_{t,other},
\end{eqnarray}
where we have made the following definitions:

\begin{eqnarray} \label{forc3}
\mathbf{F}^{Surf}_{t,kin}=-\sum_{\mathbf{k}\lambda}f^{\mathbf{k}}_{\lambda}\int_{S_{t}}d\mathbf{S}&
\Big[\Psi^{^{\dagger}\mathbf{k}(MT)}_{\lambda}(\mathbf{r})[\hat{K}-\epsilon^{\mathbf{k}}_{\lambda}]\Psi^{\mathbf{k}(MT)}_{\lambda}(\mathbf{r})\nonumber\\&-\Psi^{^{\dagger}\mathbf{k}(IR)}_{\lambda}(\mathbf{r})[\hat{K}-\epsilon^{\mathbf{k}}_{\lambda}]\Psi^{\mathbf{k}(IR)}_{\lambda}(\mathbf{r})\Big],
\end{eqnarray}

\begin{eqnarray} \label{forc4}
\mathbf{F}^{Surf}_{t,other}=&-\int_{S_{t}}d\mathbf{S}
[n^{MT}(\mathbf{r})V^{MT}_{eff}(\mathbf{r})-n^{IR}(\mathbf{r})V^{IR}_{eff}(\mathbf{r})]\nonumber\\&-\int_{S_{t}}d\mathbf{S}
[\mathbf{m}^{MT}(\mathbf{r})\cdot \mathbf{B}^{MT}_{eff}(\mathbf{r})-\mathbf{m}^{IR}(\mathbf{r})\cdot \mathbf{B}^{IR}_{eff}(\mathbf{r})]\nonumber\\&+\int_{S_{t}}d\mathbf{S}
[n^{MT}(\mathbf{r})V^{MT}_{xc}(\mathbf{r})-n^{IR}(\mathbf{r})V^{IR}_{xc}(\mathbf{r})]\nonumber\\&
+\int_{S_{t}}d\mathbf{S}
[\mathbf{m}^{MT}(\mathbf{r})\cdot \mathbf{B}^{MT}_{xc}(\mathbf{r})-\mathbf{m}^{IR}(\mathbf{r})\cdot \mathbf{B}^{IR}_{xc}(\mathbf{r})]\nonumber\\&-\int_{S_{t}}d\mathbf{S}
[n^{MT}(\mathbf{r})\epsilon^{MT}_{xc}(\mathbf{r})-n^{IR}(\mathbf{r})\epsilon^{IR}_{xc}(\mathbf{r})],
\end{eqnarray}
\section{Performance tests}
\label{conv}

\begin{table}[t]
\caption{Structural parameters of the solids considered in this work. The parameters correspond to the equilibrium geometries with zero forces. The change in atomic positions when we evaluate the forces is specified for each case later.} \label{list_s}
\begin{center}
\begin{tabular}{@{}c c c c c c c} &Space&&&&Wyckoff&$R_{MT}$\\
Solid &group&a(\AA)&b(\AA)&c(\AA)&positions&(a$_{B}$)\\
\hline\hline
$\alpha$-U&63 &2.854 &5.869 &4.955&0;0.1025;0.25  &2.602333\\
PuCoGa$_{5}$&123 &4.2354 & &6.7939&Pu: 0;0;0&Pu, Ga(1): 2.829752\\
& & & &&Co: 0;0;1/2  &Co, Ga(4): 2.34805\\
& & & &&Ga(4): 0;1/2;0.3086  &\\
& & & &&Ga(1): 1/2;1/2;0  &\\
FePt&123 &2.7248 & &3.78&Fe: 0;0;0&Fe, Pt: 2.55\\
& & & &&Pt: 1/2;1/2;1/2  &\\
\end{tabular}
\end{center}
\end{table}

\begin{table}[b]
\caption{Principal set up parameters of the studied solids.} \label{setup_s}
\begin{center}
\begin{tabular}{@{}c c c c c c} &Core&&$L_{max}$&$L_{max}$&\\
Solid &states&Semicore&$\Psi/\rho,V$&APW+lo+HDLO & $RK_{max}$ \\
\hline\hline
$\alpha$-U&[Kr]4d,4f&5s,6s,5p,6p,5d&12/8&3&12.0  \\
PuCoGa$_{5}$&Pu: [Kr]4d,4f,5s&Pu: 6s,5p,6p,5d&Pu: 12/10&Pu: 3&9.0  \\
& Co: [Ne]&Co: 3s,3p&Co: 10/10&Co: 2&  \\
& Ga: [Ne]&Ga: 3s,3p,3d&Ga: 10/10&Ga: 2&  \\
FePt&Fe: [Ne]&Fe: 3s,3p&10/10&Fe: 2&12.0  \\
& Pt: [Kr]&Pt: 5s,5p,4d,4f&&Pt: 3&  \\
\end{tabular}
\end{center}
\end{table}

This section presents results of the calculations. In order to make presentation more compact, principal structural parameters for studied solids have been collected in Table \ref{list_s} and the most important set up parameters have been collected in Table \ref{setup_s}. The APW type of the plane waves augmentation was used for the "physically relevant" orbital momenta which roughly correspond to the shells which have electrons in a free atom. This type of augmentation was accompanied with addition of two local orbitals (lo and HDLO) in order to enhance variational freedom. For higher orbital momenta, LAPW type of augmentation was applied. The separation of the augmentation strategy into APW+lo and LAPW was suggested in Ref. \cite{prb_64_195134}. Additional use of HDLO's was advocated in \cite{cpc_184_2670,cpc_220_230} and, in the context of the fully relativistic calculations, in \cite{arx_2012_04992}. High energy Local Orbitals (HELO's) were used for the "physically relevant" orbital momenta, but their effect on the calculated values of the forces was rather small. Radii of the muffin-tin spheres were selected to be the largest allowed (no overlapping). In the cases of competing sizes the ratio was 1:1. All results presented below correspond to the fully relativistic approach (FRA). A few tests performed with simplified relativistic approach (SRA, \cite{arx_2012_04992}) have shown very little difference with FRA. All calculations have been performed for the electronic temperature $T=300K$. Exchange-correlation functional corresponded to the local density approximation (LDA) as parametrized in \cite{prb_45_13244}.

\begin{table}[t]
\caption{Calculated $\alpha$-U free energy (Ry), forces (mRy/a$_{B}$), and forces evaluated by numerical differentiation of the free energy, for the different Brillouin zone samplings. Forces were evaluated for the structure with atomic positions $\mathbf{t}=\pm 0.13\mathbf{B}+1/4\mathbf{C}$ which is slightly perturbed from the equilibrium one with atomic positions $\mathbf{t}=\pm 0.1025\mathbf{B}+1/4\mathbf{C}$. The free energy corresponding to the perturbed structure is denoted as F(0) in the table. In order to evaluate the forces numerically, two additional small distortions relative to the already perturbed structure were considered. Atomic positions for these distorted structures were $\mathbf{t}=\pm (0.13\pm\Delta)\mathbf{B}+1/4\mathbf{C}$ with $\Delta=0.0005$. Free energies are given relative to the constant -112276 Ry.} \label{U_conv}
\begin{center}
\begin{tabular}{@{}c c c c c c c} Number of & &   &  & & Numerical& Mismatch    \\
\textbf{k}-points & F($-\Delta$)&   F(0)& F($+\Delta$) & Force & force &    (\%) \\
\hline\hline
144  & -0.4468887 &-0.4463283 &-0.4457661 &  -50.6287 &  -50.6095 & -0.04 \\
384  & -0.4466470 &-0.4460889 &-0.4455300 &  -50.3776 &  -50.3570 & -0.04 \\
700  & -0.4465695 &-0.4460026 &-0.4454351 &  -51.1774 &  -51.1415 & -0.07 \\
1152 & -0.4466360 &-0.4460739 &-0.4455112 &  -50.7377 &  -50.7087 & -0.06 \\
2560 & -0.4465665 &-0.4460026 &-0.4454378 &  -50.9115 &  -50.8845 & -0.05 \\
\end{tabular}
\end{center}
\end{table}

Special remark is about core states. As authors of the Ref. \cite{prb_91_035105} stress, those core states which are not exactly confined inside their MT spheres may affect the calculated forces noticeably. Such core states were allowed in \cite{prb_91_035105} to extend beyond their MT spheres and into the interstitial region and in other MT spheres with subsequent correction of the calculated forces via the plane waves expansion of their tails. This approach allows one to minimize the size of the matrices as only valence states need to be described by the basis set. The price, however, is the increased complexity of the core states treatment. Another way to handle the "shallow" core states is to include them in the list of the semicore states. In this case the size of the matrices increases slightly, but strict confinement of the remaining ('deep") core states inside their MT spheres makes the algorithm simpler, which is especially important when one builds approaches of a higher complexity (like the GW approximation) on top of the DFT code. This approach is accepted in the FlapwMBPT code.

\begin{table}[b]
\caption{Calculated free energy (Ry), forces (mRy/a$_{B}$), and numerical forces for PuCoGa$_{5}$ and ferromagnet FePt. Forces were evaluated for the structures with Pu and Pt atoms shifted from their equilibrium positions: $\mathbf{t}_{Pu}=0.02\mathbf{C}$ and $\mathbf{t}_{Pt}=1/2\mathbf{A}+1/2\mathbf{B}+0.52\mathbf{C}$ correspondingly. Free energy corresponding to these perturbed structures is denoted as F(0) in the table. In order to evaluate the forces numerically, two additional small distortions relative to already perturbed structure were considered. Plutonium positions for these distorted structures were $\mathbf{t}_{Pu}=(0.02\pm\Delta)\mathbf{C}$ with $\Delta=0.0004$, and platinum positions were: $\mathbf{t}_{Pt}=1/2\mathbf{A}+1/2\mathbf{B}+(0.52\pm\Delta)\mathbf{C}$ with $\Delta=0.0005$. Values of forces are given for Pu and Pt atoms correspondingly. Total number of \textbf{k}-points in the Brillouin zone was 486 and 6000 for PuCoGa$_{5}$ and FePt correspondingly. Free energies are given relative to the constants -81550 Ry for PuCoGa$_{5}$ and -39414 Ry for FePt.} \label{others}
\begin{center}
\begin{tabular}{@{}c c c c c c c} & &   &  & & Numerical& Mismatch    \\
Solid & F($-\Delta$)&   F(0)& F($+\Delta$) & Force & force &    (\%) \\
\hline\hline
PuCoGa$_{5}$  &-0.9305437   &-0.9303248 &-0.9301012   &-43.0625&-43.0829  &	0.05 \\
FePt  &-0.5907798  &-0.5907130 &-0.5906452  & -18.756 & -18.843	&0.46 \\
\end{tabular}
\end{center}
\end{table}

Principal results of this work, demonstrating the accuracy of the calculated forces, are collected in Table \ref{U_conv} (for $\alpha$-uranium) and in Table \ref{others} (for PuCoGa$_{5}$ and FePt). The tables also include the free energies which were used for the numerical evaluation of the forces. For the numerical differentiation we used three point formula $F'(0)=\frac{F(\Delta)-F(-\Delta)}{2\Delta}$ with $\Delta$ specified in Tables \ref{U_conv} and \ref{others}. Let us first discuss $\alpha$-uranium. As one can see from the Table \ref{U_conv}, the deviation of the calculated forces from the numerical ones is very small (about 0.05\%), which demonstrates high accuracy of the implementation. It is interesting, that the deviation is essentially independent on the sampling of the Brillouin zone. When the number of \textbf{k}-points increases, the forces and the numerical forces change slightly, but their difference is almost constant. This fact supports the robustness of the implementation. One has to mention that the forces evaluated by numerical differentiation are not exact. Not only they depend on the step $\Delta$ in the above formula (though this dependence was rather small in all cases considered in this work), but the free energies corresponding to the shift by $+\Delta$ and $-\Delta$ are subjects to different numerical errors. For instance, MT radii can be dependent on $\Delta$ (as they were in this work). Thus, comparison of the directly and numerically evaluated forces should not be considered as a test of the directly evaluated forces against the numerical ones but, rather, as a test of the consistency of the algorithms involved in both, energies and forces.

\begin{table}[t]
\caption{Calculated components of the forces exerted on all atoms. Forces correspond to the structure distorted from the equilibrium as described in tables \ref{U_conv} and \ref{others}. Surface (kinetic) term corresponds to the contribution to the force from the discontinuity of the kinetic energy at the MT surface as it is specified in the eq. (\ref{forc3}). Surface (other) include the contributions from all other discontinuities as it is specified in the eq. (\ref{forc4}). Group of four Gallium atoms (in undisturbed structure) becomes splitted in two groups (2 atoms each) with slightly different forces which are separated by the slash in the table.} \label{compon}
\begin{center}
\begin{tabular}{@{}c c c c c c c c}
Structure &$\alpha$-U &\multicolumn{4}{c}{PuCoGa$_{5}$}    &\multicolumn{2}{c}{FePt} \\
Atom & U&Pu&Co & Ga(4) & Ga(1) &Fe &Pt \\
\hline\hline
Hellmann-Feynman  &399.284   &-996.116 &-89.242   &-381.61/24.874&67.072  &-20.333	&28.551 \\
Pulay(core)  &-440.654   &954.318 & 64.177  &242.442/-2.174&-38.228  &28.868	&-46.512 \\
Pulay(valence)  &-8.712   &-9.682 &29.136   &134.506/-0.482&-27.138  &25.793	&-14.979 \\
Surface(kinetic)  &-0.883   &8.311 &-2.408   &14.172/-12.201&0.734  &-15.713	&14.208 \\
Surface(other)  &0.054   &-0.006 &-0.0004   &0.0005/-0.0003&-0.0009  &0.086	&-0.024 \\
Total  &-50.912   &-43.063 &1.663   &9.51/10.016&2.44  &18.702	&-18.756 \\
Sum of totals (drift)  & 0  &\multicolumn{4}{c}{-0.020} &\multicolumn{2}{c}{-0.054} \\
\end{tabular}
\end{center}
\end{table}

Whereas the accuracy of the basis set used in the calculations for $\alpha$-U was specially studied in our previous work \cite{arx_2012_04992}, basis sets used in the calculations for PuCoGa$_{5}$ and FePt have not been specifically tuned to reach very high accuracy. This, most likely, explains slightly bigger mismatch between the directly and numerically evaluated forces in these two cases. Nevertheless, the mismatch is small (about 0.5\%) and acceptable in most situations. It demonstrates, that the algorithm of the force evaluation is accurate enough not only when one sort of atoms is present ($\alpha$-U) but also in materials with different atoms (PuCoGa$_{5}$) and in materials with a long range magnetic order (FePt).

Finally, table \ref{compon} presents the components of the forces for all solids studied in the work. First interesting observation is that Hellmann-Feynman and Pulay (core part) are far the biggest components (especially for actinide atoms) and they cancel each other in considerable degree. Both of them come from the inner part of the MT spheres stressing the importance of correct numerical description in that area of the unit cell. Second observation is that the kinetic surface term prevails (considerably) over all other surface terms. This fact essentially supports the approximation accepted in the Ref. \cite{prb_43_6411} where only kinetic operator discontinuity was taken into account. Careful analysis of all other discontinuities performed by authors of Ref. \cite{prb_91_035105} had shown, however, the importance of these additional terms in enhancing the accuracy of the calculated forces. Thus, the other surface contributions were kept in this work and, as one can see, they are not negligible despite their relative smallness.

\section*{Conclusions}
\label{concl}

In conclusion, a formulation of the atomic forces evaluation in the framework of the relativistic density functional theory was given. It is formulated for the APW/LAPW family of basis sets with a flexible inclusion of different kind of the local orbitals (lo, HDLO, HELO). The method has been implemented in the computer code FlapwMBPT and successfully applied to the atomic forces evaluation in $\alpha$-U, PuCoGa$_{5}$, and FePt. The formulation of the forces evaluation  in the fully relativistic framework brings in an opportunity to study, for instance, the phonon spectra in actinide materials with greater reliability than it was previously available with scalar-relativistic approaches. It can also increase the efficiency of the calculations. For example, recent successful study of the phonon spectra in $\alpha$-Plutonium \cite{sr_9_18682} used the small-displacement method \cite{cpc_180_2622} and numerical differentiation of the total energies for the force evaluation. The study could be done easier with the direct evaluation of the forces.

\section*{Acknowledgments}
\label{acknow}
This work was   supported by the U.S. Department of energy, Office of Science, Basic
Energy Sciences as a part of the Computational Materials Science Program.

\bibliographystyle{elsarticle-num}

\end{document}